\documentclass[10pt, prd, twocolumn, showpacs]{revtex4}

\usepackage{amsmath}
\usepackage{amssymb}
\usepackage{graphicx}
\usepackage{bm}

\begin{document}

\title{Entropy of extremal black holes: horizon limits through charged thin
shells, a unified approach}
\author{Jos\'{e} P. S. Lemos}
\email{joselemos@ist.utl.pt}
\affiliation{Centro Multidisciplinar de Astrof\'{\i}sica, CENTRA, 
Departamento de F\'{\i}%
sica, Instituto Superior T\'ecnico - IST, Universidade de Lisboa - UL,
Avenida Rovisco Pais 1, 1049-001 Lisboa, Portugal\,\,\,}
\author{Gon\c{c}alo M. Quinta}
\email{goncalo.quinta@ist.utl.pt}
\affiliation{Centro Multidisciplinar de Astrof\'{\i}sica, 
CENTRA, Departamento de F\'{\i}%
sica, Instituto Superior T\'ecnico - IST, Universidade de Lisboa - UL,
Avenida Rovis co Pais 1, 1049-001 Lisboa, Portugal\,\,}
\author{Oleg B. Zaslavskii}
\email{zaslav@ukr.net}
\affiliation{Department of Physics and Technology, Kharkov 
V. N. Karazin National
University, 4 Svoboda Square, Kharkov 61022, Ukraine, and Institute of
Mathematics and Mechanics, Kazan Federal University, 18 Kremlyovskaya St.,
Kazan 420008, Russia}

\begin{abstract}
Using a unified approach we study the entropy of extremal black holes
through the entropy of an electrically charged thin shell. We
encounter three cases in which a shell can be taken to its own
gravitational or horizon radius and become an extremal spacetime. In
case 1, we use a non extremal shell, calculate all the thermodynamics
quantities including the entropy, take it to the horizon radius, and
then take the extremal limit. In case 2, we take the extremal limit
and the horizon radius limit simultaneously, i.e., as the shell
approaches its horizon radius it also approaches extremality. In case
3, we build first an extremal shell, and then take its horizon
radius. We find that the thermodynamic quantities in general have
different expressions in the three different cases. The entropy is the
Bekenstein-Hawking entropy $ S=A_+/4$ (where $A_+$ is the horizon
area) in cases 1 and 2, and in case 3 it can be any well-behaved
function of $A_+$. The contributions from the various thermodynamic
quantities for the entropy in all three cases are distinct. Indeed, in
cases 1 and 2 the limits agree in what concerns the entropy but they
disagree in the behavior of all other thermodynamic quantities.  Cases
2 and 3 disagree in what concerns the entropy but agree in the
behavior of the local temperature and electric potential. Case 2 is in
a sense intermediate between cases 1 and 3. Our approach sheds light
on the extremal black hole entropy issue.
\end{abstract}

\keywords{quasi-black holes, black holes, wormholes one two three}
\pacs{04.70.Bw, 04.20.Gz}
\maketitle

%\title{Entropy of an electrically charged black hole through the entropy of
%an electrically charged thin shell: extremal black holes, a
%unified approach}
%%\title{Entropy of extremal black holes: a
%unified approach through charged thin shells}

%Use showkeys class option if keyword display desired
%It is always \today, today, but any date may be explicitly specified
%\newpage

\section{Introduction}

The fact that black holes possess thermodynamic properties \cite
{hawk1,beken,hawk2} is arguably their brightest feature. Especially
fascinating is that black holes have entropy. For the nonextremal
black holes, it is known that its entropy $S$ is the
Bekenstein-Hawking entropy, equal to $A_{+}/4$, where $A_{+}$ is the
horizon area. This has been put in firm ground in the works of York
and collaborators \cite{york1,yorkmat,yorketal,93}, (see also a
generalization in \cite{can}), in a Hamiltonian formalism
\cite{lw,diaslemos}, and using quite generic matter fields \cite
{lemoszaslavskii}, among other approaches. A special kind of matter
field, thin shells, have also been used in \cite
{Mart,1,lemosquinta2,lemosminamitsuji} to further probe the
thermodynamic properties of black holes. In particular, in \cite{1}
(see also \cite {lemosquinta2,lemosminamitsuji}), the results are
based on the fact that a general thin shell can be taken to its
gravitational radius where one must force its temperature to be equal
to the Hawking temperature of a black hole, otherwise back-reaction
effects will destroy the shell. By doing so, the shell is seen to
possess an entropy equal to the Bekenstein-Hawking entropy $S=A_{+}/4$
of the correspondent spacetime black hole, thus making it possible to
calculate the entropy of an extremal black hole by using an extremal
shell taken to its gravitational radius. Notwithstanding several
efforts, it is still unclear what is microscopic explanation of this
value in the framework of a still lacking full quantum gravity.

Extremal black holes seem to be a different object  
from non-extremal ones.
Indeed, for extremal black holes, not only
the microscopic explanation of the entropy $S$ is absent, but even the value $
S$ itself of the entropy is uncertain. Although some suggestions have been
worked out that yield $S=0$ \cite{hawk3,Teit,gibbskallosh}, 
the entropy of an extremal
black hole is still an open problem, as string theory claims that it is in
fact given by the Bekenstein-Hawking entropy $S=A_+/4$ \cite
{stromingervafa,sen}, see also \cite
{ghoshmitra1995,zaslavskii1,ghoshmitra1996,ghoshmitra1997,
zaslavskii1b,zaslavskii5,mannsoloduk,
zaslavskii2,mitra1998,kieferlouko,wangsu,wangabdallasu,hod,cjr,
ederyconstantineau} on this discussion.

In Pretorius, Vollock and Israel \cite{pret} and in
\cite{lemoszaslaextremal} using matter fields, and in \cite{2} using
an extremal charged thin matter shell, an interesting solution to the
debate was naturally deduced. It was found that the extremal black
hole entropy could be any well-behaved function of $A_+$, $S=0$ and
$S=A_+/4$ included.  Of course, one might also obtain the entropy of
an extremal black hole by first calculating the entropy of a
non-extremal charged thin shell \cite{1} and then taking the extremal
limit as a particular case, this giving, as expected, $S=A_+/4$.
There is even another case, an intermediate one, when one takes the
extremal limit and the horizon radius limit simultaneously, i.e., as
the shell approaches its horizon radius it also approaches
extremality.  Therefore, it is particularly important to study the
consistency of the thin shell approach in the various limits, to
further strengthen the conclusions drawn in \cite{2}. We use the
results stated in \cite{paul1,paul2} that the thermal stress energy
tensor corresponding to a given temperature diverges in the horizon
limit unless the temperature is the Hawking temperature.

Thin shells are systems of great interest that have been used in a
number of ways in classical general relativity, as a way to quantize
gravitational systems, and concomitantly 
in a black hole context.  Classically, we
mention a variational principle found for dust shells \cite{glad}, and
the collapse of electrically charged thin shells to probe spacetime
features and test cosmic censorship 
\cite{gaolemos}. 
It has also been further used to understand
in different ways the entropy of gravitational
systems including black holes 
\cite{dav,his}.
Quantically, thin shells have for instance been used
in the understanding of quantum black hole states and Hawking
radiation, see, e.g., \cite{berezin,neronov,hajicek}.

The work is organized as follows. In Sec.~\ref{prel} we give the
preliminaries necessary to discuss the various horizon limits. We display
the first law of thermodynamics and give the expressions for the
thermodynamic quantities that enter into it. In Sec.~\ref{vari} we define
the two variables that are important to take the horizon limits, $
\varepsilon $ and $\delta $. In Sec.~\ref{geo} we define,
through geometry, the three cases
that appear in taken the horizon limit. In Sec.~\ref{mq}
we see the expressions for the mass and electric charge in the three cases.
In Sec.~\ref{pphit} we find the expressions for the 
surface pressure, the electric
potential, and the temperature in the three cases. In Sec.~\ref{s} we put
everything together into the first law and find the entropy in the three
horizon limits. In Sec.~\ref{sdisc} we discuss the contribution of each
thermodynamic quantity to the entropy and summarize these results in a
table. In Sec.~\ref{back} a discussion on the back reaction issue is raised.
In Sec.~\ref{conc} we conclude.

\section{Preliminaries}

\label{prel}

The study of the non-extremal charged thin shell developed in \cite{1}
involves three dynamical variables: the radius $R$ of the shell, its rest
mass $M$ and its charge $Q$. For thermodynamics we also need the 
local temperature $T$,
the surface pressure $p$, and the electric potential $\Phi $, and then
find the entropy $S$. Assuming that the shell is static,
spherically symmetric and has a well defined
temperature, the first law of thermodynamics is 
\begin{equation}
TdS=dM+pdA-\Phi dQ\,.  \label{1LT}
\end{equation}
where in all calculations we use natural units, the speed of light, the
gravitational constant, the Planck constant, and the Boltzmann constant are
set to one, $c=G=h=k_{B}=1$, respectively.

There are two other particularly useful variables which can characterize the
problem, namely the shell's radius $R$, the gravitational or horizon radius $
r_{+}$ and its Cauchy radius $r_{-}$, which are functions of $(R,M,Q)$
through 
\begin{equation}
r_{+}(R,M,Q)=\,m+\sqrt{m^{2}-Q^{2}}\,,  \label{horradi}
\end{equation}
\begin{equation}
r_{-}(R,M,Q)=\,m-\sqrt{m^{2}-Q^{2}}\,,  \label{horradicauch}
\end{equation}
where $m$ is the ADM mass, which can shown to be given by 
\begin{equation}
m(R,M,Q)=M-\frac{GM^{2}}{2R}+\frac{Q^{2}}{2R}\,.  \label{m}
\end{equation}
It is quite interesting that the formula given in Eq.~(\ref{m}) can be
obtained from quite different perspectives.  In \cite{yorketal}, it
was obtained from the action formalism approach to black hole
thermodynamics but it has another meaning there since it applies to
black holes, not to shells. In \cite{93}, it was rederived for bounded
self-gravitating systems using the quasilocal energy formalism.  In
\cite{1}, probably for the first time, it was obtained (i) in a pure
thermodynamic context, (ii) for thin shells, and (iii) using so
general assumptions as the first law of thermodynamics and
integrability conditions only.  In \cite{berezin}, it had been derived
from shell's dynamics.

Thus, inversely, the quantities $M$ and $Q$ can be written in terms of $
(R,r_{+},r_{-})$. Define, $k$ as 
\begin{equation}
k(R,r_{+},r_{-})=\sqrt{\Big(1-\frac{r_{+}}{R}\Big)\Big(1-\frac{r_{-}}{R}\Big)
}\,,  \label{red}
\end{equation}
usually called the redshift function. Then $M$ is given by 
\begin{equation}
M(R,r_{+},r_{-})=R(1-k)\,,  \label{M}
\end{equation}
where we have chosen the solution that gives $M=m$ for $R$ large.
Also 
\begin{equation}
Q(R,r_{+},r_{-})=\sqrt{r_{+}\,r_{-}}\,,  \label{Q}
\end{equation}
The area of the shell is 
\begin{equation}
A(R,r_{+},r_{-})=4\pi R^{2}\,,  \label{areashell}
\end{equation}
and the gravitational area or horizon area is 
\begin{equation}
A_{+}(R,r_{+},r_{-})=4\pi r_{+}^{2}\,,  \label{areahorizon}
\end{equation}
We have written explicitly the complete functional dependence $
(R,r_{+},r_{-}) $ even though some quantities do not depend on one or two of
these variables in order to show that this is a thermodynamics system. Thus, $
Q(R,r_{+},r_{-}) $ only depends on $(r_{+},r_{-})$, $A(R,r_{+},r_{-})$ only
depends on $(R)$, and $A_{+}(R,r_{+},r_{-})$ only depends on $(r_{+})$. It
will prove useful to keep the generic functional dependence.

In order for the non-extremal electric charged shell to remain static, its
surface pressure must have a specific functional form, given by \cite{1} 
\begin{equation}
p(R,r_{+},r_{-})=\frac{R^{2}(1-k)^{2}-r_{+}r_{-}}{16\pi R^{3}k}\,.
\label{pr}
\end{equation}

The electric potential $\Phi $ of the shell must also assume a specific form
if the shell is to remain static. The integrability conditions out of first
law of thermodynamics assert that \cite{1} 
\begin{equation}
\Phi(R,r_{+},r_{-})=\frac{c(r_{+},r_{-})- \frac{1}{R}}{k}\sqrt{r_{+}r_{-}}\,,
\label{phi}
\end{equation}
where $c(r_{+},r_{-})$ is as arbitrary function, which physically represents
the electric potential of the shell multiplied by its charge, if it were
located at infinity. Additionally, we need the non-extremal shell to have a
well defined electric potential in the horizon limit. This leads to 
\begin{equation}
c(r_{+},r_{-})=\frac{1}{r_{+}}\,,  \label{c}
\end{equation}
and consequently 
\begin{equation}
\Phi(R,r_{+},r_{-}) =\sqrt{\frac{r_{-}}{r_{+}}} \sqrt{\frac{1-\frac{r_{+}}{R}
}{1-\frac{r_{-} }{R}}}\,.  \label{f}
\end{equation}
Eq.~(\ref{f}) formally coincides with the expression (4.15) of \cite
{yorketal} derived for a black hole in a cavity (their $\phi $ coincides
with our $\Phi $). However, in our case there is no black hole at all.
Should the condition (\ref{c}) on the function $c$ be relaxed to an
arbitrary function, we would obtain $\lim_{R\rightarrow r_{+}}\Phi
(R,r_{+},r_{-})=\infty $, since the infinities inside the square root in the
defining Eq.~(\ref{phi}) would not be canceled.

Assuming that the shell has a well defined temperature, the integrability
conditions imposed from the first law of thermodynamics Eq.~(\ref{1LT})
gives \cite{1} 
\begin{equation}
T(R,r_{+},r_{-})=\frac{T_{0}}{k}\,,  \label{t0}
\end{equation}
where $T$ is the temperature at the shell and $T_{0}$ is the temperature
seen from infinity.

Now, we impose 
\begin{equation}
T_{0}=T_{H}(r_{+},r_{-})=\frac{r_{+}-r_{-}}{4\pi r_{+}^{2}}\,,  \label{thaw}
\end{equation}
where $T_{H}$ is the Hawking temperature of an electrically charged black
hole. So $T(R,r_{+},r_{-})=\frac{T_{H}(r_{+},r_{-})}{k}$, i.e., 
\begin{equation}
T(R,r_{+},r_{-})=\frac{r_{+}-r_{-}}{4\pi r_{+}^{2}\,k}\,.  \label{tkhaw}
\end{equation}

\section{Approach to the extremal horizon: The variables that define the
three extremal horizon limits}

\label{vari}

To study independently the limit of an extremal shell and the limit of a
shell being taken to its gravitational radius, it will prove fruitful to
define the variables $\varepsilon$ and $\delta$ through the equations 
\begin{equation}
1-\frac{r_{+}}{R}=\varepsilon ^{2}\,,  \label{e}
\end{equation}
\begin{equation}
1-\frac{r_{-}}{R}=\delta ^{2}\,.  \label{d}
\end{equation}
It is clearly seen from Eqs.~(\ref{e}) and (\ref{d}) that the variables $
\varepsilon$ and $\delta$ are the good ones to take the extremal limit.
There are however different extremal limits depending on how $\varepsilon$
and $\delta$ are taken to zero.

\section{Geometry: The three extremal horizon limits}

\label{geo}

There are three physically relevant, limits. Let us see them first through
the geometry.

\vskip 0.3cm \noindent \textbf{Case 1.} In this case we do $r_+\neq r_-$ and 
$R\to r_+$, i.e., 
\begin{equation}
\delta ={O}(1)\,,\quad\varepsilon \to 0\,.  \label{de1}
\end{equation}
After all the calculations are done and finished and we have an expression
for the entropy we can then take the $\delta\to0$ limit to get at the
gravitational radius an extremal shell. According to Eq.~(\ref{e}), this
means to bring the shell to its gravitational radius. It follows from (\ref
{d}) that $r_{+}\neq r_{-}$. Thus there is the horizon limit, but there is
no extremal limit, the shell remains nonextremal during the whole process.

\vskip 0.3cm \noindent \textbf{Case 2.} In this case we do $R\rightarrow
r_{+}$ and $r_{+}\rightarrow r_{-}$, i.e., \noindent 
\begin{equation}
\delta =\frac{\varepsilon }{\lambda}\,,\quad\varepsilon \to 0\,,  \label{de}
\end{equation}
where it is assumed that the new parameter $\lambda$ remains constant in the
limiting process and that it must satisfy $\lambda \leq 1$ due to $r_{+}\geq
r_{-}$. The limit in which $\varepsilon \rightarrow 0$, means that
simultaneously $R\rightarrow r_{+}$ and $r_{+}\rightarrow r_{-}$ in such a
way that $\delta \sim \varepsilon $. In other words, the horizon limit is
accompanied with the extremal one .

\vskip 0.3cm \noindent \textbf{Case 3.} In this case we do $r_+=r_-$ and $
R\to r_+$, i.e., 
\begin{equation}
\delta=\varepsilon \,,\quad\varepsilon \to 0\,.  \label{de2}
\end{equation}
Then, $r_{+}=r_{-}$ from the very beginning. This corresponds to the
extremal shell. This case was analyzed in \cite{2}, so we will simply state
the results and use them for comparison.

\section{Mass and electric charge: The three extremal horizon limits}

\label{mq}

Using Eqs.~(\ref{e}) and (\ref{d}) in Eq.~(\ref{red}), we immediately get
that the redshift function is 
\begin{equation}
k(R,\varepsilon,\delta)=\varepsilon \delta\,.  \label{ked}
\end{equation}
In these variables it depends on $\varepsilon$ and $\delta$ and not on $R$.

Moreover, we immediately see that 
\begin{equation}
M(R,\varepsilon,\delta)=R(1-\varepsilon \delta )\,,  \label{med}
\end{equation}
\begin{equation}
Q(R,\varepsilon,\delta)=R\sqrt{(1-\varepsilon ^{2})(1-\delta ^{2})}\,.
\label{qed}
\end{equation}
Then we can study the three cases.

\vskip 0.3cm \noindent \textbf{Case 1.} For $r_+\neq r_-$ and as $
R\rightarrow r_{+}$, i.e., for $\delta ={O}(1)$ and as $\varepsilon \to 0$,
we get from Eqs.~(\ref{med})-(\ref{ked}) 
\begin{equation}
M(r_+,\varepsilon,\delta)=r_+\,,\quad Q(r_+,\varepsilon,\delta)=r_+ \,.
\label{mqk1}
\end{equation}

\vskip 0.3cm \noindent \textbf{Case 2.} For $R\rightarrow r_{+}$ and $
r_{+}\rightarrow r_{-}$, i.e., for $\delta =\frac{\varepsilon }{\lambda} $,
with $\lambda$ kept fixed according to Eq.~(\ref{de}), and $\varepsilon \to
0 $ we get from Eqs.~(\ref{med})-(\ref{ked}) 
\begin{equation}
M(r_+,\varepsilon,\delta)=r_+\,,\quad Q(r_+,\varepsilon,\delta)=r_+ \,.
\label{mqk2}
\end{equation}

\vskip 0.3cm \noindent \textbf{Case 3.} For $r_+=r_-$ and as $R\to r_+$,
i.e., for $\delta=\varepsilon$ and $\varepsilon \to 0$ it is seen from Eq.~(
\ref{pd}) that 
\begin{equation}
M(r_+,\varepsilon,\delta)=r_+\,,\quad Q(r_+,\varepsilon,\delta)=r_+ \,.
\label{mqk3}
\end{equation}

The three limits here not surprisingly yield the same result, the
mass-charge-radius extremal condition.

\section{Pressure, electric potential and temperature: The three extremal
horizon limits}

\label{pphit}

\subsection{Pressure limits}

In order for the non-extremal electric charged shell to remain static, its
surface pressure must have a specific functional form, given by 
Eq.~(\ref{pr}) in terms of the variables $\varepsilon $ and $\delta$ defined in 
Eqs.~(\ref{e}) and (\ref{d}) can be readily written as 
\begin{equation}
p(R,\varepsilon,\delta) =\frac{1}{16\pi R}\frac{(\delta -\varepsilon )^{2}}{
\delta \varepsilon }\,.  \label{pd}
\end{equation}
Now, we will consider the behavior of pressure in all three cases.

\vskip 0.3cm \noindent \textbf{Case 1.} For $r_+\neq r_-$ and as $
R\rightarrow r_{+}$, i.e., for $\delta ={O}(1)$ and as $\varepsilon \to 0$,
we get from Eq.~(\ref{pd}) 
\begin{equation}
p(r_+,\varepsilon,\delta) = \frac{\delta }{16\pi r_+\varepsilon }\sim \frac{1
}{ \varepsilon }\,.  \label{pdiv}
\end{equation}
So, the pressure is divergent in this case as $1/\varepsilon$.

\vskip 0.3cm \noindent \textbf{Case 2.} For $R\rightarrow r_{+}$ and $
r_{+}\rightarrow r_{-}$, i.e., for $\delta =\frac{\varepsilon }{\lambda} $,
with $\lambda$ kept fixed according to Eq.~(\ref{de}), and $\varepsilon \to
0 $ we get from Eq.~(\ref{pd}) Put back intermediate step 
\begin{equation}
p(r_+,\varepsilon,\delta) =\frac{1}{ 16\pi r_{+}}\frac{(1-\lambda )^{2}}{
\lambda }\,,  \label{p3}
\end{equation}
where $\lambda=\varepsilon/\delta$ as defined in Eq.~(\ref{de}). Eq.~(\ref
{p3}) means that the pressure will remain finite but nonzero in this horizon
limit for the extremal shell.

\vskip 0.3cm \noindent \textbf{Case 3.} For $r_+=r_-$ and as $R\to r_+$,
i.e., for $\delta=\varepsilon$ and $\varepsilon \to 0$ it is seen from Eq.~(
\ref{pd}) that 
\begin{equation}
p(r_+,\varepsilon,\delta)=0\,.  \label{pe}
\end{equation}
The result $p=0$ holds in fact at any radius, including the horizon limit.

\subsection{Electric potential limits}

The electric potential $\Phi $ of the shell must also assume a specific form
if the shell is to remain static. In terms of $\varepsilon $ and $\delta $
defined in Eqs.~(\ref{e}) and (\ref{d}), Eq.~(\ref{f}) gives 
\begin{equation}
\Phi (R,\varepsilon ,\delta )=\sqrt{\frac{1-\delta ^{2}}{1-\varepsilon ^{2}}}
\,\,\frac{\varepsilon }{\delta }\,.  \label{cd}
\end{equation}
It is now straightforward to analyze the three limiting cases under
discussion.

\vskip 0.3cm \noindent \textbf{Case 1.} For $r_+\neq r_-$ and as $
R\rightarrow r_{+}$, i.e., for $\delta ={O}(1)$ and as $\varepsilon \to 0$,
we get from Eq.~(\ref{cd}), 
\begin{equation}
\Phi(r_+,\varepsilon,\delta)=0\,.
\end{equation}

\vskip 0.3cm \noindent \textbf{Case 2.} For $R\rightarrow r_{+}$ and $
r_{+}\rightarrow r_{-}$, i.e., for $\delta =\frac{\varepsilon }{\lambda}$,
with $\lambda$ kept fixed according to Eq.~(\ref{de}), and $\varepsilon \to
0 $ we get, Put back intermediate step from Eq.~(\ref{cd}), 
\begin{equation}
\Phi(r_+,\varepsilon,\delta)=\lambda\,,
\end{equation}
with $0\leq\lambda \leq 1$.

\vskip0.3cm \noindent \textbf{Case 3.} For $r_{+}=r_{-}$ and as $
R\rightarrow r_{+}$, i.e., for $\delta =\varepsilon $ and $\varepsilon
\rightarrow 0$ it would seem from Eq.~(\ref{cd}) that $\Phi
(r_{+},\varepsilon ,\delta )=1$. However, this case is special since from
the very beginning we should proceed in a different way, so the form of the
integrability condition (\ref{phi}) and Eq.~(\ref{cd}) are no longer 
valid here.
As it is shown in \cite{2}, the calculations for this case lead to 
the inequality 
\begin{equation}
\Phi (r_{+},\varepsilon ,\delta )\leq 1\,.  \label{f2}
\end{equation}
Thus, if we take an extremal shell from the very beginning, the electric
potential in general differs from what is obtained by the extremal limit
from the nonextremal state.

\subsection{Temperature limits}

Assuming that the shell has a well defined temperature, the integrability
conditions imposed from the first law of thermodynamics an in terms of $
\varepsilon$ and $\delta $ defined in Eqs.~(\ref{e}) and (\ref{d}), Eq.~(\ref
{t0}) gives 
\begin{equation}
T_{H}(R,\varepsilon,\delta)=\frac{\delta ^{2}- \varepsilon ^{2}}{ 4\pi
R(1-\varepsilon ^{2})^{2}}\,,
\end{equation}
and so the local temperature on the shell is thus 
\begin{equation}
T(R,\varepsilon,\delta)= \frac{T_{H}}{k}=\frac{\delta ^{2}-\varepsilon ^{2}}{
4\pi R\delta \varepsilon (1-\varepsilon ^{2})^{2}}\,.  \label{tloc}
\end{equation}

\vskip 0.3cm \noindent \textbf{Case 1.} For $r_+\neq r_-$ and as $
R\rightarrow r_{+}$, i.e., for $\delta ={O}(1)$ and as $\varepsilon \to 0$,
we get from Eq.~(\ref{tloc}), 
\begin{equation}  \label{tdiv}
T(r_+,\varepsilon,\delta)= \frac{\delta }{4\pi r_{+}\varepsilon }\sim
\frac1\varepsilon \,.
\end{equation}
So it diverges.

\vskip 0.3cm \noindent \textbf{Case 2.} For $R\rightarrow r_{+}$ and $
r_{+}\rightarrow r_{-}$, i.e., for $\delta =\frac{\varepsilon }{\lambda}$,
with $\lambda$ kept fixed according to Eq.~(\ref{de}), and $\varepsilon \to
0 $ we get from Eq.~(\ref{tloc}), Put back intermediate step 
\begin{equation}
T(r_+,\varepsilon,\delta) =\frac{1-\lambda ^{2}}{4\pi r_{+}\lambda }\,.
\label{t3}
\end{equation}
It remains finite and nonzero. It is worth noting a simple formula that
follows from (\ref{t3}) and relates the pressure and temperature in this
horizon limit, namely, $\frac{p}{T}=\frac{1}{4}\frac{1-\lambda }{1+\lambda }
. $

\vskip 0.3cm \noindent \textbf{Case 3.} For $r_+=r_-$ and as $R\to r_+$,
i.e., for $\delta=\varepsilon$ and $\varepsilon \to 0$, one can relax
condition (\ref{thaw}) in such a way that $T_{0}\rightarrow 0$ but $T$
remains finite (see \cite{2} fo details).

\section{Entropy: The three extremal horizon limits}
\label{s}

To obtain the distinct limits for the entropy, one can express the first law
of thermodynamics Eq.~(\ref{1LT}) in terms of the variables $(R,r_+,r_-)$,
using the Eqs.~(\ref{M}), (\ref{Q}), (\ref{areashell}), (\ref{pr}), (\ref
{phi}), and (\ref{tkhaw})). In turn using Eqs.~(\ref{e}), (\ref{d}), (\ref
{ked}), (\ref{med}), (\ref{qed}), (\ref{pd}), and (\ref{cd}), the first law
of thermodynamics Eq.~(\ref{1LT}) can be expressed in terms of the variables 
$(R,\varepsilon,\delta)$, in the quite general exact form $
TdS=a_{1}dR+a_{2}d\varepsilon +a_{3}d\delta \label{1LTd} $, where $
a_{1}=1-\delta \varepsilon +\frac{(\delta -\varepsilon )^{2}}{2\delta
\varepsilon }+\frac{(1-\delta ^{2})(1-\varepsilon ^{2})}{\delta \epsilon }
(1-Rc) $, $a_{2}=-\delta R\left[ 1+\frac{1-\delta ^{2}}{\delta ^{2}}(1-Rc)
\right] $, $a_{3}=-\varepsilon R\left[ 1+\frac{1-\varepsilon ^{2}}{
\varepsilon ^{2}} (1-Rc)\right] $. Imposing further that the electric
potential must also assume the value of Eq.~(\ref{c}), enables to simplify
the coefficients $a_1$, $a_2$, and $a_3$, into $a_{1}=\frac{\delta
^{2}-\varepsilon ^{2}}{2\delta \varepsilon } $, $a_{2}=-\delta R\left[ 1-
\frac{\varepsilon ^{2}}{\delta ^{2}}\left( \frac{ 1-\delta ^{2}}{
1-\varepsilon ^{2}}\right) \right] $, $a_{3}=0 $. Then, using Eq.~(\ref{tloc}
), the differential for the entropy in the variables $(R,\epsilon,\delta)$
becomes 
\begin{equation}
dS(R,\epsilon,\delta)=2\pi \,R \left(1-\varepsilon^2\right)^2\,dR
-4\pi\,R^2\varepsilon\left(1-\varepsilon^2\right)d\varepsilon\,.  \label{ds}
\end{equation}
This equation can be integrated to give 
\begin{equation}
S(r_+,\epsilon,\delta)=\pi \,R^2 \left(1-\varepsilon^2\right)^2\,,
\label{ds2}
\end{equation}
where we have put the integration constant to zero. Using Eq.~(\ref{e}) it
gives 
\begin{equation}
S(r_+)=\frac{A_+}{4}\,  \label{sbh00}
\end{equation}
where $A_+$ is the gravitational radius area, or the horizon area when the
shell is push into the gravitational radius, see Eq.~(\ref{areahorizon}). It
is the Bekenstein-Hawking entropy. It is striking that all the other
quantities, $p$, $\Phi$, $T$, depend generically on $\varepsilon$ and $
\delta $. The entropy does not, it only depends on $r_+$.

\vskip 0.3cm \noindent \textbf{Case 1.} For $r_+\neq r_-$ and as $
R\rightarrow r_{+}$, i.e., for $\delta ={O}(1)$ and as $\varepsilon \to 0$,
we get from Eq.~(\ref{sbh00}), $S(r_+)=\frac{A_+}{4}$. This is general for
any nonextremal black hole. We can now take the extremal limit $\delta\to0$
and obtain that the entropy of an extremal charged black hole is by
continuity $S(r_+)=\frac{A_+}{4}$, the Bekenstein-Hawking entropy.

\vskip 0.3cm \noindent \textbf{Case 2.} For $R\rightarrow r_{+}$ and $
r_{+}\rightarrow r_{-}$, i.e., for $\delta =\frac{\varepsilon }{\lambda}$,
with $\lambda$ kept fixed according to Eq.~(\ref{de}), and $\varepsilon \to
0 $ we obtain from Eq.~(\ref{sbh00}), $S(r_+)=\frac{A_+}{4}$. So in the case
that the shell achieves the gravitational radius simultaneously with the
extremal limit one also gets the Bekenstein-Hawking entropy.

\vskip 0.3cm \noindent \textbf{Case 3.} For $r_+=r_-$ and as $R\to r_+$,
i.e., for $\delta=\varepsilon$ and $\varepsilon \to 0$, the entropy cannot
be handled in this manner and should be considered separately. This has been
done in \cite{2} with the result that the entropy is not fixed unambiguously
for a given $r_{+}$, it is any physical well behaved function of $r_+$, or
if one prefers, of $A_+$, i.e., 
\begin{equation}
S(r_+)= \mathrm{a\;physical\;well\;behaved\;function\;of\;}A_+\,.
\label{sbh3}
\end{equation}

Eqs.~(\ref{ds})-(\ref{sbh00}) work for cases 1 and 2. In case 3, the ab
initio extremal shell with $\delta =\varepsilon$ one is led to the
discussion given in \cite{2}.

\section{Discussion on the three extremal horizon limits: Where does the
entropy stem from?}

\label{sdisc}

It is instructive to trace in more detail, how from the first law, the
entropy arises. More precisely, we are interested in the question: Which
contributions dominate for the three different cases?

\vskip 0.3cm \noindent \textbf{Case 1.} For $r_+\neq r_-$ and as $
R\rightarrow r_{+}$, i.e., for $\delta ={O}(1)$ and as $\varepsilon \to 0$,
let us, for simplcity, do $\varepsilon =\mathrm{constant}\ll 1$. Then, in
the first law Eq.~(\ref{1LT}), and from Eq.~(\ref{pdiv}), we can retain the
term due to the pressure only, taking into account Eq.~(\ref{tdiv}), we
obtain the result (\ref{sbh00}). Thus, the pressure term gives the whole
contribution to the entropy. See also \cite{lemoszaslavskii}.

\vskip 0.3cm \noindent \textbf{Case 2.} For $R\rightarrow r_{+}$ and $
r_{+}\rightarrow r_{-}$, i.e., for $\delta =\frac{\varepsilon }{\lambda}$,
with $\lambda$ kept fixed according to Eq.~(\ref{de}), and $\varepsilon \to
0 $, all three terms in the first law give contribution to the entropy.
Thus, the mass, pressure and electric potential terms give contributions to
the entropy.

\vskip 0.3cm \noindent \textbf{Case 3.} For $r_+=r_-$ and as $R\to r_+$,
i.e., for $\delta=\varepsilon$ and $\varepsilon \to 0$, and according to
Eq.~(\ref{pe}), the first and third terms in Eq.~(\ref{1LT}) contribute to
the entropy. Thus cases 1 and 3 are complementary to each other in what
concerns the origin of the entropy.

\vskip 0.3cm It is convenient to present the results in the following table.
It is implied that in all three cases the horizon limit is taken.

\begin{widetext}
\hskip -0.5cm
\begin{tabular}
[c]{|l|l|l|l|l|l|}\hline
Case & Pressure $p$ & Potential $\Phi$ & Local temperature $T$ & Entropy &
Contribution from (according to 1st law)\\\hline
1 & divergent like $\varepsilon^{-1}$ & 1 & infinite & $A_+/4$ &
pressure\\\hline
2 & finite nonzero & any$<1$ & finite nonzero & $A_+/4$ & mass, pressure and
potential\\\hline
3 & 0 & any$\leq1$ & finite nonzero & a function of $A_+$& 
mass and potential\\\hline
\end{tabular}
\vskip 0.2cm
\noindent
Table 1. The contributions of the pressure
$p$, electric potential $\Phi$, and temperature $T$,
to the 
extremal black hole entropy $S$, according to the first law.
\label{tabent}
\end{widetext}

It is worth stressing that the results presented in the table refer in
general not to black holes but to shells. Only in the horizon limit these
results apply to black holes. Usually, if one considers the extremal limit
of a nonextremal black hole, it remains in the same topological class during
the limiting transition, so it is not surprising that in the extremal limit
one obtains the Bekenstein-Hawking value. However, in our case, we obtained
something more: the fact that the exact value of the shell's entropy coincides
with that of a black hole for a given $r_{+}$ independently of $R$. For an
arbitrary self-gravitating matter system this is not so, 
the entropy of the system 
is a function of $r_+$, $R$, and possibly other variables. 
Only in the horizon
limit the Bekenstein-Hawking value is recovered \cite{lemoszaslavskii}.

\section{Role of the backreaction}

\label{back}

As is known, for a nonextremal spacetime, the thermal stress energy tensor
corresponding to a temperature $T_{0}$ can be represented in the form \cite
{paul1,paul2} 
\begin{equation}
T_{\mu }^{\nu }=\frac{T_{0}^{4}-T_{H}^{4}}{(g_{00})^{2}} f_{\mu }^{\nu }\,,
\end{equation}
where, $f_{\mu }^{\nu }$ is some tensor finite on the horizon, $g_{00}$
being the $00$ component of the metric in use. In the horizon limit, the
requirement of the finiteness of $T_{\mu }^{\nu }$ entails $T_{0}=T_{H}$.
For the nonextremal horizon one has $T_{H}\neq 0$.

Now, in the extremal case, $T_{H}=0$, whence 
\begin{equation}
T_{\mu }^{\nu }=\frac{T_{0}^{4}}{(g_{00})^{2}} f_{\mu }^{\nu }\,.
\end{equation}
Thus, the attempt to put $T_{0}\neq 0$ according to the prescriptions given
in \cite{hawk3,Teit}, leads to infinite stresses since $\frac{T_{0}^{4}}{
(g_{00})^{2}}$ diverges as one approaches the horizon. This destroys the
horizon \cite{paul1,paul2}.

However, when we deal with a shell instead of a black hole, an intermediate
case can be realized. Namely, simultaneously $T_{0}\rightarrow 0 $ and $
g_{00}\rightarrow 0$ in such a way that $T$ is kept bounded. This is
realized in Case 2 according to Eq.~(\ref{t3}). It is also realized in Case
3.

\section{Conclusions}

\label{conc}

We found what happens 
in calculating the entropy and the other thermodynamic
quantities 
when different limiting transitions for a shell are taken
and how they are related to each other when the
radius of the shell approaches the horizon radius, 
i.e., it turns into a black hole.

It happens that the limits in cases 
1 and 2 agree in what concerns the entropy but
they disagree in the behavior of all other quantities. Cases 2 and 3
disagree in what concerns the entropy but agree in the behavior of the
local temperature and electric potential. Case 2 is 
intermediate between 1 and 3.

The results obtained showed how careful one should be in the
calculations when a system approaches the horizon which, in turn, is
close to the extremal state. It is of interest to trace whether and
how these subtleties can affect calculations in quantum field theory
including string theory.

\begin{acknowledgments}
We thank FCT-Portugal for financial support through Project
No.~PEst-OE/FIS/UI0099/2015. GQ 
also ack\-now\-ledges the grant No.~SFRH/BD/92583/2013 from FCT.
OBZ has been partially supported by the Kazan Federal University 
through a state grant for scientific activities.
\end{acknowledgments}

\end{document}